\documentclass[twocolumn,showpacs,amsmath,amssymb,10pt,aps]{revtex4}
\usepackage{graphicx,color}
\usepackage{amsmath}
\usepackage{amssymb}
\usepackage{bbm, bm}
\usepackage{color}

\usepackage[T1]{fontenc}
\usepackage[cp1250]{inputenc}
\usepackage{amsfonts}
\usepackage{amssymb, amsbsy}
\usepackage{amsmath, bbm}
\usepackage{epstopdf}

\begin{document}
\title{Quantum trajectories for a system interacting with environment in a single photon state: counting and diffusive processes}
\author{Anita D\k{a}browska$^1$, Gniewomir Sarbicki$^2$, and Dariusz Chru\'sci\'nski$^2$}
\affiliation{ $^1$Nicolaus Copernicus University, Toru\'{n}, Collegium Medicum Bydgoszcz, ul. Jagiello\'{n}ska 15, 85-067 Bydgoszcz, Poland\\
$^2$Institute of Physics, Faculty of Physics, Astronomy and Informatics,  Nicolaus Copernicus University,
Grudzi\c{a}dzka 5/7, 87--100 Toru\'n, Poland}

\begin{abstract}

We derived quantum trajectories for a system interacting with the environment prepared in a continuous mode single photon state as the limit of discrete filtering model with an environment defined as series of independent qubits prepared initially in the entangled state being an analogue of a continuous mode state. The environment qubits interact with the quantum system and they are subsequently measured. The initial correlation between the bath qubits is the source of the non-Markovianity. The conditional evolutions of the quantum system for limit of the continuous in time observations together with the formulas for the photon counting probabilities are given.

\end{abstract}

\maketitle

\section{Introduction}

Belavkin filtering equations \cite{BB91,BGM04, B05} (called also stochastic master equations \cite{Car93, BP02, GZ10, WM10}) describe the conditional evolution of an open quantum system interacting with the environment modeled by the Bose field \cite{Bar06}. The quantum filtering theory, providing the best state estimation based
on the result of the continuous in time measurement, is formulated within the framework of quantum stochastic It\^{o} calculus (QSC) \cite{HP84, Par92}. The stochastic evolution of the system depends on the results of the measurement performed continuously in time on the output field - the field after interaction with the system. The indirect measurement allows us to avoid Zeno effect. The exact form of the filtering equation depends on the initial state of the environment and the measurement scheme. The rigorous derivation of the filtering equation for the Bose field in a Gaussian state (vacuum, coherent, squeezed) one can find for instance in \cite{BB91, BGM04, DG16}. Solutions to the filtering equations are called quantum trajectories. 

Recently there is interest in determination of the stochastic master equation for non-classical states of the Bose field. One of them is a single photon state \cite{L00,RMS07, M08}. About the methods of its generation and application in quantum of computing and communication one can read in \cite{LHABMS01, RNBW05, SAL10a, SAL10b, BRHS10, WMSS11, BCBC12, BAM12}. The filtering equation for quantum system interacting with the environment prepared in the single photon state was derived in \cite{GJN11, GJNC12a, GJN12b, GJN13}. The authors of the mentioned papers determined the filters with the aid of embedding of the quantum system into a larger Markovian \cite{GJN11, GJNC12a} or non-Markovian \cite{GJN12b, GJN13} systems. In \cite{GJN11, GJNC12a} the authors used the concept of quantum cascaded system \cite{GZ10} and introduced an ancilla being a source of the Bose field in desired non-classical state. The output field from ancilla is the input field for the quantum system. The authors first derived the filters for the extended system consisting of the system of interest and ancilla and then, by taking the partial trace over the ancilla degrees of freedom, they determined the conditional evolution for the quantum system. The cascaded systems approach to determine the conditional evolution of a quantum system interacting with the single photon, which one can find for instance in \cite{BAM12}, was first proposed in the paper \cite{GEPZ98}. Derivation of the quantum trajectories for the light prepared in a continuous-mode Fock state in the framework of QSC was given in \cite{SZX16, BC17}.  

In the paper we present new derivation of the filters for quantum system interacting with an environment in the single photon state as a limit of discrete filtering. To determine the quantum trajectories we use, instead of the standard methods based on the concept of ancilla producing the field in non-classical state, the quantum repeated interaction model \cite{AP06, P08, PP09, P10}. We present the model with a quantum system interacting with an infinite chain of identical and independent quantum systems representing an environment. This approach gives an intuitive and rigorous interpretation for the conditional evolution of the open quantum and quantum trajectories.

We consider the model with an environment defined by a series of independent qubits.  The environment qubits interact in turn one by one with the quantum system and they are subsequently measured. Each qubit of the environment interacts with the system only once and the environment qubits do not interact between themselves. We assume that the environment is prepared initially in an entangled state being a discrete analogue of a continuous mode single-photon state. We obtain the stochastic
evolutions of the quantum system for the two cases of the monitoring of the environment qubits: photodetection and homodyne detection. The conditional evolutions of the system are derived by making use of the von Neumnann projection postulate. We showed that the quantum system becomes entangled with this part of the environment which has not interacted with the quantum system yet. The prior and posterior evolution of the quantum system are non-Markovian in this case. The memory effects result from the initial correlation between the bath qubits. Together with the stochastic evolution we present also the formulas for the photon counting probabilities giving the whole statistics of the output photons.

The model is based on the discrete version of quantum It\^{o} stochastic calculus.
Its rigorous definition together with discussion of its continuous limit
were given in number of publications \cite{M93, AB93, G04, GS04, P05, S05, AP06, K06, BH08, BHJ09}. Derivations of discrete version of quantum filtering equations with its continuous limits for the case when the environment is initially in factorized state with the qubits prepared in some mixed state one can find in \cite{B02} and for the qubits prepared in the ground state (the environment prepared in the vacuum state) in \cite{GS04, BHJ09}.

\section{Repeated quantum interactions model}

Let us consider a quantum system $\mathcal{S}$ of the Hilbert space $\mathcal{H}_{\mathcal{S}}$ interacting with the environment consisting of a sequence of two-level systems which interact in turn one by one with the system $\mathcal{S}$ each during the time interval of the length $\tau$.
Thus Hilbert space of environment is
\begin{equation}
\mathcal{H}_{\mathcal{E}}=\bigotimes_{k=0}^{+\infty}\mathcal{H}_{\mathcal{E},k},
\end{equation}
where $\mathcal{H}_{\mathcal{E},k}=\mathbb{C}^2$ is the Hilbert space of the $k$-th qubit interacting with $\mathcal{S}$ in the time interval $[k\tau, (k+1)\tau)$. The Hilbert space $\mathcal{H}_{\mathcal{E}}$ can be split as a tensor product
\begin{equation}
\mathcal{H}_{\mathcal{E}}=\mathcal{H}_{\mathcal{E}}^{j-1]}\otimes \mathcal{H}_{\mathcal{E}}^{[j}\,,\;\;\;\mathcal{H}_{\mathcal{E}}^{j-1]}=\bigotimes_{k=0}^{j-1}\mathcal{H}_{\mathcal{E},k}\,,\;\;\;
\mathcal{H}_{\mathcal{E}}^{[j}=
\bigotimes_{k=j}^{+\infty}\mathcal{H}_{\mathcal{E},k}.
\end{equation}
If $j\tau$ is the current moment then $\mathcal{H}_{\mathcal{E}}^{j-1]}$ can be interpreted as the part of the space of the environment which refer to two-level systems which have already interacted with $\mathcal{S}$ and $\mathcal{H}_{\mathcal{E}}^{[j}$ as the space referring to two-level systems which have not interacted with $\mathcal{S}$ yet. We will call them respectively the past and future environment spaces.

The ground and excited states of the $k$-th two-level system we will indicate respectively by $|0\rangle_{k}$ and $|1\rangle_{k}$. We assume that the environment is prepared in the state
\begin{eqnarray}
|1_{\xi}\rangle=  \sqrt{\tau} \sum_{k=0}^{+\infty}\xi_{k}\sigma^{+}_{k}|vac\rangle,
\end{eqnarray}
where
$|vac\rangle=|0\rangle_{0}\otimes |0\rangle_{1} \otimes |0\rangle_{2} \otimes |0\rangle_{3} \ldots$
is the vacuum vector in $\mathcal{H}_{\mathcal{E}}$, the operators $\sigma^{-}_{k}=|0\rangle_{k}\langle 1|$, $\sigma^{+}_{k}=|1\rangle_{k}\langle 0|$ act non-trivially only in the space $\mathcal{H}_{\mathcal{E},k}$, and $\displaystyle{\sum_{k=0}^{+\infty}}|\xi_{k}|^2\tau = 1$. Note that $|1_{\xi}\rangle$ has the additive decomposition property
\begin{equation}
|1_{\xi}\rangle
=  \sqrt{\tau} \sum_{k=0}^{j}\xi_{k}\sigma^{+}_{k}|vac\rangle +   \sqrt{\tau} \sum_{k=j+1}^{+\infty}\xi_{k}\sigma^{+}_{k}|vac\rangle
\end{equation}
and it can be written in the form
\begin{equation}
|1_{\xi}\rangle =   \sqrt{\tau}  \sum_{k=0}^{+\infty}\xi_{k}|1_k \rangle,
\end{equation}
where
\begin{equation}
|1_k \rangle=|0\rangle_{0}\otimes |0\rangle_{1} \otimes\ldots |0\rangle_{k-1}\otimes|1\rangle_{k}\otimes |0\rangle_{k+1}\otimes |0\rangle_{k+2}\ldots.
\end{equation}
Thus $|\xi_{k}|^2\tau$ is the probability that $k$-th qubit is prepared in the upper state and all the others qubits are in their ground states. The state $|1_{\xi}\rangle$ is a discrete analogue of a single-photon state in the model of continuous modes discussed in \cite{RMS07, M08}.

Now, let us describe the evolution of the composed $\mathcal{E}+\mathcal{S}$ system. We consider the well known repeated interactions model where the unitary operator defining the evolution of the total system  up to the time $j\tau$ is given by
\begin{equation}
U_{j\tau} = \mathbb{V}_{j-1} \mathbb{V}_{j-2} \ldots \mathbb{V}_{0},\;\;\;\;\;U_{0}=\mathbbm{1},
\end{equation}
where $\mathbb{V}_{k}$ acts non-trivially only in the space $\mathcal{H}_{\mathcal{E},k}\otimes \mathcal{H}_{\mathcal{S}}$ and
\begin{equation}
\mathbb{V}_{k}=\exp\left(-i \tau H_{k}\right),
\end{equation}
where $H_k$ is a bipartite Hamiltonian acting on $\mathcal{H}_{\mathcal{E},k} \otimes  \mathcal{H}_S$. It is therefore clear that there is no interaction between subsystems $\mathcal{H}_{\mathcal{E},k}$ and  $\mathcal{H}_{\mathcal{E},k'}$ for $k \neq k'$.
 Since $\mathcal{H}_{\mathcal{E},k} = \mathbb{C}^2$ one has the following representation
\begin{equation}
  \mathbb{V}_k = \left( \begin{array}{cc}    V_{00} & V_{01} \\ V_{10} & V_{11} \end{array} \right) ,
\end{equation}
with $V_{ij}$ being the system operators. We assume that the initial state of the total system is the pure product state of the form
\begin{equation}
|1_{\xi}\rangle\otimes|\psi\rangle
\end{equation}
and we stress that the environmental state $|1_{\xi}\rangle$ is highly entangled (depending on the profile $\xi_{k}$).

\section{Conditional evolution for the counting process}

In what follows we consider the following collision model: the system $\mathcal{S}$ interacts with an infinite chain of environmental qubits. Now, after each interaction one performs a  measurement on the last qubit which has just interacted with $\mathcal{S}$. Our goal is to describe an evolution of $\mathcal{S}$ conditioned on the results of the measurements performed subsequently on the environment qubits at the time instances $\tau, 2\tau, 3\tau, \ldots$. Clearly, since the results of measurements are random it gives rise to a random evolution of the state of $\mathcal{S}$. 

Let us first consider the measurement of the observable
\begin{equation}\label{observable1}
\sigma_{k}^{+}\sigma_{k}^{-}=|1\rangle_{k}\langle 1|,\;\;k=0, 1, 2, \ldots.
\end{equation}
We shall prove  that the conditional state of $\mathcal{S}$ and the part of the environment which has not interacted with $\mathcal{S}$ up to $j\tau $ is at the moment $j\tau $ given by
\begin{equation}\label{cond2}
|\tilde{\Psi}_{j| \pmb{\eta}_j}\rangle = \frac{|\Psi_{j| \pmb{ \eta}_j}\rangle}{\sqrt{\langle\Psi_{j| \pmb{\eta}_j}|\Psi_{j| \pmb{\eta}_j}\rangle}},
\end{equation}
where $ |\Psi_{j|\pmb{\eta}_j} \rangle $ is the unnormalized conditional vector of the form
\begin{eqnarray}\label{M}
  |\Psi_{j|\pmb{\eta}_j} \rangle &=&  \sqrt{\tau} \sum_{k=j}^{+\infty} \xi_{k} \sigma_{k}^{+} | vac \rangle_{[j,+\infty)} \otimes |\alpha_{j| \pmb{\eta}_j}\rangle \nonumber \\ &+& | vac \rangle_{[j,+\infty)} \otimes |\beta_{j| \pmb{\eta}_j}\rangle
\end{eqnarray}
where $\pmb{\eta}_j$ is a binary $j$-vector $\pmb{\eta}_j = (\eta_j,\eta_{j-1},\ldots,\eta_1)$ with $\eta_k \in \{0,1\}$ which represents results of all measurements of (\ref{observable1}) up to the time $j\tau$. We use an obvious notation
\begin{equation}
|vac\rangle_{[j,+\infty)}=|0\rangle_{j}\otimes |0\rangle_{j+1} \otimes\ldots.
\end{equation}
The vectors $|\alpha_{j| \pmb{\eta}_j}\rangle$, $|\beta_{j| \pmb{\eta}_j}\rangle$ from the Hilbert space $\mathcal{H}_{\mathcal{S}}$ satisfy the recurrence equations
\begin{eqnarray}\label{rec1a}
    |\alpha_{j+1| \pmb{\eta}_{j+1} }\rangle &=& V_{\eta_{j+1} 0} |\alpha_{j| \pmb{\eta}_j}\rangle,\\
\label{rec1b}
    |\beta_{j+1| \pmb{\eta}_{j+1} }\rangle &=& V_{\eta_{j+1} 0} |\beta_{j| \pmb{\eta}_j}\rangle + \sqrt{\tau} \xi_{j} V_{\eta_{j+1} 1} |\alpha_{j| \pmb{\eta}_j}\rangle 
  \end{eqnarray}
with the initial condition $|\alpha_0\rangle=|\psi\rangle$, $|\beta_0\rangle=0$.

Note that the conditional vectors $|\alpha_{j|  \pmb{\eta}_j}\rangle$, $|\beta_{j|  \pmb{\eta}_j}\rangle$  depend on all results of the measurements up to the time $j\tau$. Their physical interpretation becomes clear when we write down explicitly their form for particular trajectories. Usually, one skips the condition $\pmb{\eta}_j$. In this Section, however, to stress the character of the conditional states, we keep the conditional notation. 

 It is clear that
$ |\widetilde{\Psi}_{j|\pmb{\eta}_j} \rangle$ belongs to the Hilbert space $\displaystyle{\bigotimes_{k=j}^{+\infty}}\mathcal{H}_{\mathcal{E},k}\otimes \mathcal{H}_{\mathcal{S}}$. The form of $|\Psi_{j|  \pmb{\eta}_{j} }\rangle$ indicates that the system $\mathcal{S}$ becomes entangled with this part of the environment which has not interacted with $\mathcal{S}$ yet. It is the main difference between the considered situation and the standard cases when the environment is in the factorisable state \cite{GS04, BHJ09}. The physical interpretation of  $|\Psi_{j|  \pmb{\eta}_{j} }\rangle$ is very intuitive:

\begin{itemize}
\item the first term in (\ref{M}) represents the following scenario: all qubits of the environment up to time $j\tau$ were prepared in the ground state and the qubit prepared in the excited state appears only in the future,
     
\item the second term represents the situation in which  $\mathcal{S}$ has already interacted with the qubit prepared in the excited state and in the future it will interact with the environment being in the vacuum. 
\end{itemize}
These two scenarios occur as a quantum superposition. The probability of the first one is
\begin{equation}
\frac{\langle\alpha_{j|  \pmb{\eta}_{j}}|\alpha_{j|  \pmb{\eta}_{j}}\rangle \sum_{k=j}^{+\infty}\tau|\xi_{k}|^2}{\langle\alpha_{j|  \pmb{\eta}_{j}}|\alpha_{j|  \pmb{\eta}_{j}}\rangle \sum_{k=j}^{+\infty}\tau|\xi_{k}|^2+\langle\beta_{j|  \pmb{\eta}_{j}}|\beta_{j|  \pmb{\eta}_{j}}\rangle}
\end{equation}
and the probability of the second one is
\begin{equation}
\frac{\langle\beta_{j|  \pmb{\eta}_{j}}|\beta_{j|  \pmb{\eta}_{j}}\rangle}{\langle\alpha_{j|  \pmb{\eta}_{j}}|\alpha_{j|  \pmb{\eta}_{j}}\rangle \sum_{k=j}^{+\infty}\tau|\xi_{k}|^2+\langle\beta_{j|  \pmb{\eta}_{j}}|\beta_{j|  \pmb{\eta}_{j}}\rangle}.
\end{equation}
Sooner or later, it depends in the profile $\xi_{k}$ of the environment state $|1_{\xi}\rangle$, the system $\mathcal{S}$ meets the qubit prepared in the excited state, so finally only the second term gives non-zero contribution to (\ref{M}), and $|\Psi_{j|  \pmb{\eta}_{j} }\rangle$ becomes separable.

In order to prove of (\ref{M}) it is convenient to rewrite $|\Psi_{j|  \pmb{\eta}_j} \rangle$ in the following form
\begin{eqnarray}
    |\Psi_{j|  \pmb{\eta}_j} \rangle &=& |0\rangle_{j} \otimes \Big( \sum_{k=j+1}^{+\infty} \sqrt{\tau}\xi_k\sigma_{k}^{+} |vac \rangle_{[j+1,+\infty)} \otimes |\alpha_{j|  \pmb{\eta}_j}\rangle  \nonumber \\ &+&  | vac \rangle_{[j+1,+\infty)} \otimes |\beta_{j|  \pmb{\eta}_j}\rangle \Big)\nonumber \\
&+& |1\rangle_{j}\otimes |vac\rangle_{[j+1,+\infty)}\otimes \sqrt{\tau}\xi_{j}|\alpha_{j|  \pmb{\eta}_j}\rangle.
  \end{eqnarray}
Now the action of $\mathbb{V}_{j}$ on $|\Psi_{j|\pmb{\eta}_j} \rangle$ gives
\begin{widetext}
\begin{eqnarray}\label{cond}
&& \mathbb{V}_{j} |\Psi_{j| \pmb{\eta}_j} \rangle = |0\rangle_{j} \otimes\bigg(\sum_{k=j+1}^\infty \sqrt{\tau}\xi_{k} \sigma_{k}^{+}|vac \rangle_{[j+1,+\infty)} \otimes V_{00}|\alpha_{j|  \pmb{\eta}_j} \rangle + | vac \rangle_{[j+1,+\infty)} \otimes
\left( V_{00}|\beta_{j|  \pmb{\eta}_j} \rangle + \sqrt{\tau}\xi_{j}V_{01}|\alpha_{j|  \pmb{\eta}_j} \rangle \right)\bigg)  \nonumber\\ && + |1\rangle_{j} \otimes\bigg(\sum_{k=j+1}^\infty \sqrt{\tau}\xi_{k} \sigma_{k}^{+}| vac\rangle_{[j+1,+\infty)} \otimes V_{10}|\alpha_{j|  \pmb{\eta}_j} \rangle + | vac \rangle_{[j+1,+\infty)} \otimes \left(V_{10}|\beta_{j|  \pmb{\eta}_j} \rangle+ \sqrt{\tau}\xi_{j}V_{11}|\alpha_{j|  \pmb{\eta}_j} \rangle\right)\bigg).
\end{eqnarray}
\end{widetext}
The conditional vector $|\Psi_{j+1|  \pmb{\eta}_{j+1}} \rangle$ from the Hilbert space   $\displaystyle{\bigotimes_{k=j+1}^{+\infty}}\mathcal{H}_{\mathcal{E},k}\otimes \mathcal{H}_{S}$ is defined by
\begin{equation}
\left(\Pi_{\eta_{j+1}}\otimes \mathbbm{1}_{S}\right) \mathbb{V}_{j}|\Psi_{j|  \pmb{\eta}_j} \rangle=|\eta_{j+1}\rangle_{j}\otimes|\Psi_{j+1|  \pmb{\eta}_{j+1} }\rangle,
\end{equation}
where
\begin{eqnarray}
\Pi_{0}=|0\rangle_{j}\langle 0|,\;\;\;\Pi_{1}=|1\rangle_{j}\langle 1|,
\end{eqnarray}
and $\eta_{j+1}=0,1$  stands for the two possible results of the measurement of the observable $\sigma_{j}^{+}\sigma_{j}^{-}$. Finally,  using (\ref{cond}) one finds that $|\Psi_{j+1|  \pmb{\eta}_{j+1} }\rangle$ has the following form
\begin{eqnarray}\label{cond1}
   |\Psi_{j+1|  \pmb{\eta}_{j+1} }\rangle &=& \sum_{k=j+1}^{+\infty} \xi_{k} \sqrt{\tau}\sigma_{k}^{+} | vac \rangle_{[j+1,+\infty)} \otimes |\alpha_{j+1| \pmb{\eta}_{j+1} }\rangle \nonumber\\ &+& | vac \rangle_{[j+1,+\infty)} \otimes |\beta_{j+1|  \pmb{\eta}_{j+1} }\rangle
  \end{eqnarray}
with  $|\alpha_{j+1|  \pmb{\eta}_j+1}\rangle$, $|\beta_{j+1|  \pmb{\eta}_j+1}\rangle$  satisfying Eqs. (\ref{rec1a}) and (\ref{rec1b}),which ends the proof of (\ref{M}).

Note that knowing the initial state $|\psi\rangle$ of $\mathcal{S}$, single photon state $|1_\xi\rangle$ of the environment, and the results of measurements $(\eta_j,\ldots,\eta_1)$ up to time $j \tau$ one can uniquely determine the state $|\Psi_j|\pmb{\eta}_j\rangle$.

Formula (\ref{cond}) shows that after each measurement there are in general two possible scenarios: 

\begin{itemize}

\item the system $\mathcal{S}$ meets in the future the qubit prepared in the excited state,
 
\item or it meets in the future all qubits prepared in the ground state. 
\end{itemize}
In the quantum filtering theory the environment is usually treated as a model of electromagnetic field. In this interpretation we see in (\ref{cond}) that when the result of the outcome is $0$ then we deal with two possibilities---the system $\mathcal{S}$ did not emit photon or photon from the environment was absorbed by $\mathcal{S}$. When the photon was detected we see  that it was emitted by $\mathcal{S}$ or we measured the photon of the external field directly. The form of (\ref{M}) depends on the results of all past measurements, thus it is clear that if all photons achievable for the considered composed system were observed then only the second term gives non-zero contribution to (\ref{M}) and there is no longer entanglement between $\mathcal{S}$ and the future environment.

The general solution to the set of equations (\ref{rec1a}) and  (\ref{rec1b}) can be written in the compact form
  \begin{eqnarray}
    |\alpha_{j|\pmb{\eta}_{j}}\rangle = V_{\eta_{j} 0} V_{\eta_{j-1} 0} \dots V_{\eta_1 0} |\psi\rangle, \\
    |\beta_{j|\pmb{\eta}_{j}}\rangle = \sqrt{\tau} \sum_{k=0}^{j-1}  \xi_{k}\prod_{l=1}^{j} V_{\eta_l \delta_{l(k+1)}} |\psi\rangle.
  \end{eqnarray}
Note that instead of using the notation with the full vector $\pmb{\eta}_j$ we may equivalently provide the location of `$1$' in the string $(\eta_j,\ldots,\eta_1)$, that is $(l_m,\ldots,l_1)$ means that one observed $m$ photons at $\tau_i = \tau l_i$ $(i=1,\ldots,m)$ and no other photons in the period from $0$ to $j\tau$. Let us provide the formulae for the simplest scenarios:

\begin{enumerate}

\item $\pmb{\eta}_j=(0,\ldots,0) = \mathbf{0}_j$, that is, no detections upon $\tau j$. In this case one finds

\begin{equation}\label{convec1a}
|\alpha_{j|\mathbf{0}_j}\rangle=V_{00}^{j}|\psi\rangle,
\end{equation}
\begin{equation}\label{convec1b}
|\beta_{j|\mathbf{0}_j}\rangle= \sqrt{\tau} \sum_{k=0}^{j-1}V_{00}^{j-k-1}\xi_{k}V_{01}V_{00}^k|\psi\rangle.
\end{equation}

\item one detection at $\tau_1 = \tau l_1$:

\begin{equation}\label{convec2a}
|\alpha_{j|l_1}\rangle=
V_{00}^{j-l_{1}}V_{10}V_{00}^{l_{1}-1}|\psi\rangle,
\end{equation}
and
\begin{eqnarray}\label{convec2b}
|\beta_{j|l_1}\rangle&=& \sqrt{\tau}
\left[V_{00}^{j-l_{1}}\xi_{l_{1}-1}V_{11}V_{00}^{l_{1}-1} \right. \nonumber \\ &+&
V_{00}^{j-l_{1}}V_{10}\sum_{k=0}^{l_{1}-2}V_{00}^{l_{1}-2-k}\xi_{k}V_{01}V_{00}^k \\
&+ &\left. \sum_{k=l_{1}}^{j-1}V_{00}^{j-k-1}\xi_{k}V_{01}V_{00}^{k-l_{1}}V_{10}V_{00}^{l_{1}-1}
\right]|\psi\rangle. \nonumber
\end{eqnarray}

\item two detections $(l_2,l_1)$:

\begin{equation}\label{convec3a}
|\alpha_{j|l_{2},l_{1}}\rangle=
V_{00}^{j-l_{2}}V_{10}V_{00}^{l_{2}-l_{1}-1}V_{10}V_{00}^{l_{1}-1}|\psi\rangle,
\end{equation}
and
\begin{eqnarray}\label{convec3b}
&& |\beta_{j|l_{2},l_{1}}\rangle = \sqrt{\tau}
\left[V_{00}^{j-l_{2}}V_{10}V_{00}^{l_{2}-l_{1}-1}\xi_{l_{1}-1}V_{11}V_{00}^{l_{1}-1}\right.\nonumber\\
&&+V_{00}^{j-l_{2}}\xi_{l_{2}-1}V_{11}V_{00}^{l_{2}-l_{1}-1}V_{10}V_{00}^{l_{1}-1} \\
&&+V_{00}^{j-l_{2}}V_{10}V_{00}^{l_{2}-l_{1}-1}V_{10}\sum_{k=0}^{l_{1}-2}V_{00}^{l_{1}-k-2}\xi_{k}V_{01}V_{00}^k\nonumber\\
&&+V_{00}^{j-l_{2}}V_{10}\sum_{k=l_{1}}^{l_{2}-2}V_{00}^{l_{2}-k-2}\xi_{k}V_{01}V_{00}^{k-l_{1}}V_{10}V_{00}^{l_{1}-1}\nonumber\\
&&\left.+\sum_{k=l_{2}}^{j-1}V_{00}^{j-k-1}\xi_{k}V_{01}V_{00}^{k-l_{2}}V_{10}V_{00}^{l_{2}-l_{1}-1}V_{10}V_{00}^{l_{1}-1}
\right]|\psi\rangle. \nonumber
\end{eqnarray}

\end{enumerate}

In the representation of the moments of counts $(l_m,\ldots,l_1)$ one finds
 \begin{eqnarray}
    |\alpha_{j|l_m,\dots,l_1}\rangle = V_{00}^{j}\prod_{q=1}^{\stackrel{\longleftarrow}{m}}V_{10}^{(l_{q})}|\psi\rangle,
  \end{eqnarray}
 and quite involved formula for $|\beta_j\rangle$:
  \begin{eqnarray}  \label{Gniewko}
&& |\beta_{j|l_m,\dots,l_1}\rangle = \sqrt{\tau}V_{00}^j\left[  \sum_{p=1}^m \prod_{q=p+1}^{\stackrel{\longleftarrow}{m}} V_{10}^{(l_q)} \xi_{l_p-1} V_{11}^{(l_p)} \prod_{q=1}^{\stackrel{\longleftarrow}{p-1}} V_{10}^{(l_q)}
     \right.\nonumber \\
     && \left.   + \sum_{p=0}^m\prod_{q=p+1}^{\stackrel{\longleftarrow}{m}} V_{10}^{(l_q)} W_{l_{p}}^{l_{p+1}}\prod_{q=1}^{\stackrel{\longleftarrow}{p}} V_{10}^{(l_q)} \right] | \psi\rangle,
  \end{eqnarray}
where we defined
$$V_{ij}^{(l)} {=} V_{00}^{-l} V_{ij} V_{00}^{l-1},$$
and

$$ W_{l_p}^{l_{p+1}} {=} \sqrt{\tau} \sum_{r=l_{p+1}}^{l_{p+1}-1}\xi_{r-1} V_{01}^{(r)}  . $$
provided $l_0 = 0$ and $l_{m+1}=j+1$. The arrows mean that we use the time ordered products.

In order to obtain the state of $\mathcal{S}$ conditioned on all results of the measurements up to the time $j\tau$, we have to perform a partial trace of $|\tilde{\Psi}_{j|  \pmb{\eta}_{j}}\rangle\langle\tilde{\Psi}_{j|  \pmb{\eta}_{j}}|$ with respect to the environment degrees of freedom (the future space of the environment). Thus the {\it a posteriori} state of $\mathcal{S}$ at the time $j\tau$ is
\begin{equation}\label{condS}
\tilde{\rho}_{j|  \pmb{\eta}_{j}}
=\frac{\rho_{j|  \pmb{\eta}_{j}}}{\mathrm{Tr}\rho_{j|  \pmb{\eta}_{j}}},
\end{equation}
where
\begin{equation}
\rho_{j|  \pmb{\eta}_{j}} = |\alpha_{j|  \pmb{\eta}_{j}}\rangle\langle\alpha_{j|  \pmb{\eta}_{j}}|\;\displaystyle{\sum_{k=j}^{+\infty}}\;\tau|\xi_{k}|^2 +
|\beta_{j|  \pmb{\eta}_{j}}\rangle\langle\beta_{j|  \pmb{\eta}_{j}}|.
\end{equation}
The probability of a particular trajectory $\pmb{\eta}_{j}$ registered from time $0$ to $j\tau$ is therefore given by
\begin{equation}\label{conprob}
\mathrm{Tr}\rho_{j|  \pmb{\eta}_{j}}=\langle\alpha_{j|  \pmb{\eta}_{j}}|\alpha_{j|  \pmb{\eta}_{j}}\rangle \sum_{k=j}^{+\infty}\tau|\xi_{k}|^2+
\langle\beta_{j|  \pmb{\eta}_{j}}|\beta_{j|  \pmb{\eta}_{j}}\rangle.
\end{equation}

\section{Discrete filtration equation for the counting process}

\subsection{General case}

From now on we simplify our notation to $|\alpha_j\rangle$ and $|\beta_j\rangle$ skipping the condition $\pmb{\eta}_j$. We derive the conditional recurrence equation describing the stochastic evolution of $\mathcal{S}$. Consider first the case when the result of measurement at $(j+1)\tau$ is $0$. Straightforward calculations lead to
\begin{eqnarray}\label{Filter1}
&& \rho_{j+1} = V_{00} \rho_{j} V_{00}^\dagger +  \sqrt{\tau} \left( \xi_j V_{01} |\alpha_j\rangle \langle \beta_j| V_{00}^\dagger + \mbox{h.c.} \right) \nonumber\\
&& + \tau |\xi_j|^2 \Big(  V_{01}  |\alpha_{j}\rangle\langle\alpha_{j}| V_{01}^\dagger -  V_{00}
 |\alpha_{j}\rangle\langle\alpha_{j}| V_{00}^\dagger \Big) ,
\end{eqnarray}
where h.c. stands for the Hermitian conjugation. Moreover, one finds
\begin{eqnarray}\label{}
  |\alpha_{j+1}\rangle\langle\beta_{j+1}| &=&  V_{00} |\alpha_{j}\rangle\langle\beta_{j}|V_{00}^\dagger + \sqrt{\tau} \xi_j^*  V_{00} |\alpha_{j}\rangle\langle\alpha_{j}|V_{01}^\dagger,\nonumber\\
|\alpha_{j+1}\rangle\langle\alpha_{j+1}| &=& V_{00} |\alpha_{j}\rangle\langle\alpha_{j}|V_{00}^\dagger.
\end{eqnarray}
The conditional probability of the outcome $\eta_{j+1}=0$  when  the {\it a posteriori} state of $\mathcal{S}$ at $j\tau$ was $\tilde{\rho}_{j}$ is given by the formula
\begin{equation}
p_{j+1}(0|\tilde{\rho}_{j})=\frac{\mathrm{Tr}\rho_{j+1|0}}{\mathrm{Tr}\rho_{j}}.
\end{equation}
Similarly, if $\eta_{j+1}=1$ one obtains

\begin{eqnarray}\label{Filter2}
&& \rho_{j+1} = V_{10} \rho_{j} V_{10}^\dagger +  \sqrt{\tau} \left( \xi_j V_{11} |\alpha_j\rangle \langle \beta_j| V_{10}^\dagger + \mbox{h.c.} \right) \nonumber\\
&& + \tau |\xi_j|^2 \Big(  V_{11}  |\alpha_{j}\rangle\langle\alpha_{j}| V_{11}^\dagger -  V_{10}
 |\alpha_{j}\rangle\langle\alpha_{j}| V_{10}^\dagger \Big)
\end{eqnarray}
and
\begin{eqnarray}
  |\alpha_{j+1}\rangle\langle\beta_{j+1}| &=&  V_{10} |\alpha_{j}\rangle\langle\beta_{j}|V_{10}^\dagger + \sqrt{\tau} \xi_j^*  V_{10} |\alpha_{j}\rangle\langle\alpha_{j}|V_{11}^\dagger,\nonumber\\
|\alpha_{j+1}\rangle\langle\alpha_{j+1}| &=& V_{10} |\alpha_{j}\rangle\langle\alpha_{j}|V_{10}^\dagger.
\end{eqnarray}
The conditional probability of the outcome $\eta_{j+1}=1$  when  the {\it a posteriori} state of $\mathcal{S}$ at $j\tau$ was $\tilde{\rho}_{j}$ is given by the formula
\begin{equation}\label{condprob}
p_{j+1}(1|\tilde{\rho}_{j})=\frac{\mathrm{Tr}\rho_{j+1|1}}{\mathrm{Tr}\rho_{j}} ,
\end{equation}
and clearly $p_{j+1}(0|\tilde{\rho}_{j})  + p_{j+1}(1|\tilde{\rho}_{j}) = 1$.

\subsection{`Small' $\tau$}

To consider the case of `small' $\tau$ one needs to specify the form of the Hamiltonian. Let us consider \cite{AP06, GS04}
\begin{eqnarray}\label{}
  H_{k} = \mathbbm{1}_{k}\otimes H_{\mathcal{S}} +\frac{i}{\sqrt{\tau}}\left(\sigma_{k}^{+}\otimes L-
\sigma_{k}^{-}\otimes L^{\dagger}\right) ,
\end{eqnarray}
which means that we work in the interaction picture eliminating the free evolution of the environment.

Now, a time step $\tau$ is small if
\begin{equation}
  \tau  \omega_{\rm max}  = \tau \max_{ij} |E_i - E_j| / \hbar  \ll 1 ,
\end{equation}
and $E_i$ belongs to the spectrum of $H_\mathcal{S}$, that is, $1/\tau$ is much larger the characteristic frequency of the system.
One easily finds
\begin{eqnarray}\label{vmatrix}
V_{00}&=& \mathbbm{1}_{\mathcal{S}} - i\tau H_{\mathcal{S}} - \tau \frac{1}{2}L^\dagger L + O(\tau^{2}) ,\nonumber \\
V_{10}&=&\sqrt{\tau} L + O(\tau^{3/2}) ,\nonumber \\
V_{01}&=&- \sqrt{\tau} L^\dagger + O(\tau^{3/2}) ,\\
V_{11}&=& \mathbbm{1}_{\mathcal{S}} + O(\tau)\nonumber  .
\end{eqnarray}

Now, discrete filtration equation  (\ref{Filter1}) gives rise to
\begin{eqnarray}\label{filter1}
\rho_{j+1} &=& \rho_{j}-i[H_{S},\rho_{j}]\tau-\frac{1}{2}\left\{L^{\dagger}L,\rho_{j}\right\}\tau\nonumber\\
&-&|\beta_{j}\rangle\langle\alpha_{j}|L\xi_{j}^{\ast}\tau-L^{\dagger}|\alpha_{j}\rangle\langle\beta_{j}|\xi_{j}\tau
\nonumber\\ &-& |\alpha_{j}\rangle\langle\alpha_{j}||\xi_{j}|^2\tau + O(\tau^2) ,
\end{eqnarray}
together with
\begin{eqnarray}
&& |\alpha_{j+1}\rangle\langle\beta_{j+1}|
= |\alpha_{j}\rangle\langle\beta_{j}| \nonumber\\
&&- i\left[H_{S},|\alpha_{j}\rangle\langle\beta_{j}|\right]\tau
-\frac{1}{2}\left\{L^{\dagger}L,|\alpha_{j}\rangle\langle\beta_{j}|\right\}\tau\nonumber\\
&&-|\alpha_{j}\rangle\langle\alpha_{j}|L\xi_{j}^{\ast}\tau + O(\tau^2) ,
\end{eqnarray}
and
\begin{eqnarray}
&& |\alpha_{j+1}\rangle\langle\alpha_{j+1}| = |\alpha_{j}\rangle\langle\alpha_{j}| \\ &&-i[H_{S},|\alpha_{j}\rangle\langle\alpha_{j}|]\tau
 -\frac{1}{2}\left\{L^{\dagger}L,|\alpha_{j}\rangle\langle\alpha_{j}|\right\}\tau + O(\tau^2). \nonumber
\end{eqnarray}
The conditional  probability $p_{j+1}(0|\tilde{\rho}_{j})$ reads
\begin{equation}
p_{j+1}(0|\tilde{\rho}_{j})=1- k_{j}\tau,
\end{equation}
where
\begin{eqnarray}\label{intensity1}
 k_{j} &=& \mathrm{Tr}\left( L^{\dagger}L\tilde{\rho}_{j}
+ \frac{\xi_{j}^{\ast}}{\mathrm{Tr}\rho_{j}}\, L|\beta_{j}\rangle\langle\alpha_{j} |
\right. \nonumber\\ &+& \left.    \frac{\xi_{j}}{\mathrm{Tr}\rho_{j}} \,
|\alpha_{j}\rangle\langle\beta_{j}|L^{\dagger}
+ \frac{|\xi_{j}|^2}{\mathrm{Tr}\rho_{j}}\,|\alpha_{j}\rangle\langle\alpha_{j}|  \right).
\end{eqnarray}
Now, using
\begin{equation}
\frac{1}{\mathrm{Tr}\rho_{j+1}} =\frac{1}{ \mathrm{Tr}\rho_j  }    \left(1+k_{j} \tau \right) + O(\tau^2) ,
\end{equation}
one obtains the following equation  for the normalized density matrix (we neglect higher order terms in $\tau$)
\begin{eqnarray}
&& \tilde{\rho}_{j+1} = \tilde{\rho}_{j} +  \tilde{\rho}_{j} k_{j}\tau-i[H_{S},\tilde{\rho}_{j}]\tau-\frac{1}{2}\left\{L^{\dagger}L, \tilde{\rho}_{j}\right\}\tau\nonumber\\
&-&|\tilde{\beta}_{j}\rangle\langle \tilde{\alpha}_{j}|L\xi_{j}^{\ast}\tau-L^{\dagger}|\tilde{\alpha}_{j}\rangle\langle\tilde{\beta}_{j}|\xi_{j}\tau
\nonumber\\&-&|\tilde{\alpha}_{j}\rangle\langle\tilde{\alpha}_{j}||\xi_{j}|^2\tau.
\end{eqnarray}
Similarly, for the conditional vectors defined by
\begin{equation}
|\tilde{\alpha}_{j}\rangle=\frac{|{\alpha}_{j}\rangle}{\sqrt{\mathrm{Tr}\rho_{j}}},\;\;\;\;\;
|\tilde{\beta}_{j}\rangle=\frac{|{\beta}_{j}\rangle}{\sqrt{\mathrm{Tr}\rho_{j}}},
\end{equation}
one finds the following formulae (provided  $\eta_{j+1}=0$):
\begin{eqnarray}
|\tilde{\alpha}_{j+1}\rangle\langle\tilde{\beta}_{j+1}|
&=&|\tilde{\alpha}_{j}\rangle\langle\tilde{\beta}_{j}|+|\tilde{\alpha}_{j}\rangle\langle\tilde{\beta}_{j}|k_{j}\tau
-i\left[H_{S},|\tilde{\alpha}_{j}\rangle\langle\tilde{\beta}_{j}|\right]\tau\nonumber\\
&-&\frac{1}{2}\left\{L^{\dagger}L,|\tilde{\alpha}_{j}\rangle\langle\tilde{\beta}_{j}|\right\}\tau
-|\tilde{\alpha}_{j}\rangle\langle\tilde{\alpha}_{j}|L\xi_{j}^{\ast}\tau,
\end{eqnarray}
and
\begin{eqnarray}
|\tilde{\alpha}_{j+1}\rangle\langle\tilde{\alpha}_{j+1}|&=&|\tilde{\alpha}_{j}\rangle\langle\tilde{\alpha}_{j}|
+|\tilde{\alpha}_{j}\rangle\langle\tilde{\alpha}_{j}|k_{j}\tau
-i[H_{S},|\tilde{\alpha}_{j}\rangle\langle\tilde{\alpha}_{j}|]\tau\nonumber\\
&-& \frac{1}{2}\left\{L^{\dagger}L,|\tilde{\alpha}_{j}\rangle\langle\tilde{\alpha}_{j}|\right\}\tau.
\end{eqnarray}
If $\eta_{j+1}=1$ discrete filtration equation  (\ref{Filter2}) gives rise to
\begin{eqnarray}\label{filter2}
\rho_{j+1}&=&L\rho_{j}L^{\dagger}\tau+L|\beta_{j}\rangle\langle\alpha_{j}|
\xi_{j}^{\ast}\tau+|\alpha_{j}\rangle\langle\beta_{j}|L^{\dagger}\xi_{j}\tau \nonumber \\ &+&
|\alpha_{j}\rangle\langle\alpha_{j}||\xi_{j}|^2\tau ,
\end{eqnarray}
\begin{eqnarray}
|\alpha_{j+1}\rangle\langle\beta_{j+1}|
&=&L|\alpha_{j}\rangle\langle\beta_{j}|L^{\dagger}\tau+L|\alpha_{j}\rangle\langle\alpha_{j}|\xi_{j}^{\ast}\tau
\end{eqnarray}
and
\begin{eqnarray}
|\alpha_{j+1}\rangle\langle\alpha_{j+1}|
&=&L|\alpha_{j}\rangle\langle\alpha_{j}|L^{\dagger}\tau.
\end{eqnarray}
The conditional probability of the outcome $1$ at the moment $(j+1)\tau$ when  the {\it a posteriori} state of $\mathcal{S}$ at $j\tau$ is $\tilde{\rho}_{j}$ has the form 
\begin{equation}
p_{j+1}(1|\tilde{\rho}_{j})=k_{j}\tau,
\end{equation}
where $k_{j}$ is given by (\ref{intensity1}).

Finally, for the normalized density matrix, we get
\begin{eqnarray}
\tilde{\rho}_{j+1}&=&
\frac{1}{k_{j}}\bigg(L\tilde{\rho}_{j}L^{\dagger}
+L|\tilde{\beta}_{j}\rangle\langle\tilde{\alpha}_{j}|\xi_{j}^{\ast}+
|\tilde{\alpha}_{j}\rangle\langle\tilde{\beta}_{j}|L^{\dagger}\xi_{j}\nonumber
\\&&+|\tilde{\alpha}_{j}\rangle\langle\tilde{\alpha}_{j}||\xi_{j}|^2\bigg),
\end{eqnarray}
\begin{eqnarray}
|\tilde{\alpha}_{j+1}\rangle\langle\tilde{\beta}_{j+1}|
\!\!&=&\!\!\frac{1}{k_{j}}\left(L|\tilde{\alpha}_{j}\rangle\langle\tilde{\beta}_{j}|L^{\dagger}+
L|\tilde{\alpha}_{j}\rangle\langle\tilde{\alpha}_{j}|\xi_{j}^{\ast}\right),
\end{eqnarray}
\begin{eqnarray}
|\tilde{\alpha}_{j+1}\rangle\langle\tilde{\alpha}_{j+1}|
&=&\frac{1}{k_{j}}L|\tilde{\alpha}_{j}\rangle\langle\tilde{\alpha}_{j}|L^{\dagger}.
\end{eqnarray}
It is clear that $\eta_k$ $(k=1,2,\ldots)$ are random variables with values $\{0,1\}$ and hence one deals with the discrete stochastic process $(\eta_j,\eta_{j-1},\ldots,\eta_{1})$. A single realization of this process consists of zeros and ones and the conditional expectations for $\eta_{j+1}$ read
\begin{eqnarray}
\mathbbm{E}[\eta_{j+1}|\tilde{\rho}_{j}]&=& k_{j}\tau + O(\tau^2) ,\nonumber\\
\mathbbm{E}[(\eta_{j+1})^2|\tilde{\rho}_{j}]&=& k_{j}\tau + O(\tau^2).
\end{eqnarray}
Introducing $  n_j = \sum_{k=1}^j \eta_k $ one has an obvious relation
$$\eta_{j+1} = n_{j+1} - n_j =: \Delta n_j. $$  
Hence $n_j$ may be interpreted as a discrete counting process. Therefore to describe the stochastic evolution of $\mathcal{S}$ depending on the (stochastic) results of the measurements we need the following set of three coupled equations:

\begin{widetext}
 \begin{eqnarray}\label{filter3}
\tilde{\rho}_{j+1}&=& \tilde{\rho}_{j}-i[H_{S},\tilde{\rho}_{j}]\tau-\frac{1}{2}\left\{L^{\dagger}L,\tilde{\rho}_{j}\right\}\tau
+L\rho_{j}L^{\dagger}\tau 
+ [|\tilde{\alpha_{j}}\rangle\langle\tilde{\beta}_{j}|,L^{\dagger}]\xi_{j}\tau
+[L, |\tilde{\beta}_{j}\rangle\langle\tilde{\alpha}_{j}|]\xi^{\ast}_{j}\tau \nonumber\\
&+& \bigg\{\frac{1}{k_{j}}\left(
L\tilde{\rho}_{j}L^{\dagger}+L|\tilde{\beta}_{j}\rangle\langle\tilde{\alpha}_{j}|\xi^{\ast}_{j}+|\tilde{\alpha}_{j}\rangle\langle\tilde{\beta}_{j}| L^{\dagger}\xi_{j}
+
|\tilde{\alpha}_{j}\rangle\langle\tilde{\alpha}_{j}||\xi_{j}|^2\right)
-\tilde{\rho}_{j}\bigg\}(\Delta n_{j+1}-k_{j}\tau),
\end{eqnarray}

\begin{eqnarray}  \label{I}
 |\tilde{\alpha}_{j+1}\rangle\langle\tilde{\beta}_{j+1}| &=& |\tilde{\alpha}_{j}\rangle\langle\tilde{\beta}_{j}|
-i\left[H_{S},|\tilde{\alpha}_{j}\rangle\langle\tilde{\beta}_{j}|\right]\tau 
-\frac{1}{2}\left\{L^{\dagger}L,|\tilde{\alpha}_{j}\rangle\langle\tilde{\beta}_{j}|\right\}\tau 
+L|\tilde{\alpha}_{j}\rangle\langle\tilde{\beta}_{j}|L^{\dagger}\tau + \left[L,|\tilde{\alpha}_{j}\rangle\langle\tilde{\alpha}_{j}|\right]\xi^{\ast}_{j}\tau\nonumber\\
&+& \bigg\{\frac{1}{k_{j}}\left(L|\tilde{\alpha}_{j}\rangle\langle\tilde{\beta}_{j}|L^{\dagger}+
L|\tilde{\alpha}_{j}\rangle\langle\tilde{\alpha}_{j}|
\xi^{\ast}_{j}\right) -|\tilde{\alpha}_{j}\rangle\langle\tilde{\beta}_{j}|\bigg\}(\Delta n_{j+1}-k_{j}\tau),
\end{eqnarray}

\begin{eqnarray}   \label{II}
|\tilde{\alpha}_{j+1}\rangle\langle\tilde{\alpha}_{j+1}|&=&
|\tilde{\alpha}_{j}\rangle\langle\tilde{\alpha}_{j}|-i\left[H_{S},|\tilde{\alpha}_{j}\rangle\langle\tilde{\alpha}_{j}|\right]\tau
-\frac{1}{2}\left\{L^{\dagger}L,|\tilde{\alpha}_{j}\rangle\langle\tilde{\alpha}_{j}|\right\}\tau
+L|\tilde{\alpha}_{j}\rangle\langle\tilde{\alpha}_{j}|L^{\dagger}\tau\nonumber\\
&+&\bigg\{\frac{1}{k_{j}}L|\tilde{\alpha}_{j}\rangle\langle\tilde{\alpha}_{j}|L^{\dagger}
-|\tilde{\alpha}_{j}\rangle\langle\tilde{\alpha}_{j}|\bigg\}(\Delta n_{j+1}-k_{j}\tau)
\end{eqnarray}
\end{widetext}
with the initial condition $\tilde{\rho}_{0}=|\psi\rangle\langle\psi|$, $|\tilde{\alpha}_{0}\rangle\langle\tilde{\beta}_{0}|=0$, and $|\tilde{\alpha}_{0}\rangle\langle\tilde{\alpha}_{0}|=|\psi\rangle\langle\psi|$. If $\Delta n_{j+1}=1$ then all terms containing the infinitesimal $\tau$ are negligible.

It should be clear that as an initial condition one may take an arbitrary mixed state $\rho_0$. Then one replaces $|\tilde{\alpha}_{j}\rangle\langle\tilde{\alpha}_{j}|$ by $X_j$ and $|\tilde{\alpha}_{j}\rangle\langle\tilde{\beta}_{j}|$ by $Y_j$ with initial conditions
$X_0 = \rho_0$ and $Y_0 = 0$.

\section{Continuous limit for the counting process}

In the continuous limit  $\tau\rightarrow 0$ one obtains from (\ref{filter3})-(\ref{II}) the stochastic differential equations
\begin{widetext}
\begin{eqnarray}\label{filcont1}
d\tilde{\rho}_{t}&=&-i[H_{S},\tilde{\rho}_{t}]dt-
\frac{1}{2}\left\{L^{\dagger}L,\tilde{\rho}_{t}\right\}dt
+L\tilde{\rho}_{t}L^{\dagger}dt +[\tilde{\rho}^{01}_{t},L^{\dagger}]\xi_{t}dt
+[L, \tilde{\rho}^{10}_{t}]\xi^{\ast}_{t}dt\nonumber\\
&+& \bigg\{\frac{1}{k_{t}}\left(
L\tilde{\rho}_{t}L^{\dagger}+L\tilde{\rho}^{10}_{t}\xi^{\ast}_{t}+\tilde{\rho}^{01}_{t}L^{\dagger}\xi_{t}
+\tilde{\rho}^{00}_{t}|\xi_{t}|^2\right)
-\tilde{\rho}_{t}\bigg\}\left(dn(t)-k_{t}dt\right),
\end{eqnarray}

\begin{eqnarray}\label{filcont2}
d\tilde{\rho}^{01}_{t}&=& -i[H_{S},\tilde{\rho}^{01}_{t}]dt-\frac{1}{2}\left\{L^{\dagger}L,\tilde{\rho}^{01}_{t}
\right\}dt+L\tilde{\rho}^{01}_{t}L^{\dagger}dt + \left[L,\tilde{\rho}^{00}_{t}\right]\xi^{\ast}_{t}dt \nonumber\\
&+& \left\{\frac{1}{k_{t}}\left(L\tilde{\rho}^{01}_{t}L^{\dagger}+
L\tilde{\rho}^{00}_{t}\xi^{\ast}_{t}\right)-\tilde{\rho}^{01}_{t}\right\}\left(dn(t)-k_{t}dt\right),
\end{eqnarray}

\begin{eqnarray}\label{filcont3}
d\tilde{\rho}^{00}_{t}&=&-i[H_{S},\tilde{\rho}^{00}_{t}]dt-\frac{1}{2}\left\{L^{\dagger}L,\tilde{\rho}^{00}_{t}
\right\}dt+L\tilde{\rho}^{00}_{t}L^{\dagger}dt + \left(\frac{1}{k_{t}}L\tilde{\rho}^{00}_{t}L^{\dagger}-
\tilde{\rho}^{00}_{t}\right)\left(dn(t)-k_{t}dt\right),
\end{eqnarray}
\end{widetext}
where $\tilde{\rho}^{10}_{t}=(\tilde{\rho}^{01}_{t})^{\dagger}$ and the initial condition $\tilde{\rho}_{0}=|\psi\rangle\langle\psi| $, $\tilde{\rho}^{01}_{0}= 0$, and $\tilde{\rho}^{00}_{0}=|\psi\rangle\langle\psi|$. The stochastic process $n(t)$ satisfies
\begin{equation}
\mathbbm{E}[dn(t)|\tilde{\rho}_{t}]=k_{t}dt
\end{equation}
and
\begin{equation}
k_{t}=\mathrm{Tr}\left(L^{\dagger}L\tilde{\rho}_{t}
+L\tilde{\rho}^{10}_{t}\xi_{t}^{\ast}+\tilde{\rho}^{01}_{t}L^{\dagger}\xi_{t}+\tilde{\rho}^{00}_{t}|\xi_{t}|^2\right).
\end{equation}
One can easily check that for the non-selective measurement we obtain from (\ref{filcont1})-(\ref{filcont3}) {\it a priori} evolution given by
 \begin{eqnarray}
\dot{\tilde{\rho}}_{t}&=&-i[H_{S},\tilde{\rho}_{t}]-
\frac{1}{2}\left\{L^{\dagger}L,\tilde{\rho}_{t}\right\}
+L\tilde{\rho}_{t}L^{\dagger}\nonumber\\&+&
 [\tilde{\rho}^{01}_{t},L^{\dagger}]\xi_{t}
+[L, \tilde{\rho}^{10}_{t}]\xi^{\ast}_{t},
\end{eqnarray}
\begin{eqnarray}
\dot{\tilde{\rho}}^{01}_{t}&=& -i[H_{S},\tilde{\rho}^{01}_{t}]-\frac{1}{2}\left\{L^{\dagger}L,\tilde{\rho}^{01}_{t}
\right\}+L\tilde{\rho}^{01}_{t}L^{\dagger}\nonumber\\
&+& \left[L,\tilde{\rho}^{00}_{t}\right]\xi^{\ast}_{t},
\end{eqnarray}
\begin{eqnarray}
\dot{\tilde{\rho}}^{00}_{t}= -i[H_{S},\tilde{\rho}^{00}_{t}]
-\frac{1}{2}\left\{L^{\dagger}L,\tilde{\rho}^{00}_{t}
\right\}+L\tilde{\rho}^{00}_{t}L^{\dagger}.
\end{eqnarray}
All realization of the counting process $n(t)$ may be divided into disjoint sectors: $\mathcal{C}_m$ contains realizations with exactly $m$ counts at some moments $t_m > \ldots > t_2 > t_1 > 0$ and no other photons from $0$ to $t$. Denote by $ p_{0}^{t}(t_{m}, t_{m-1}, \ldots, t_{2}, t_{1})$ the probability density of observing a particular trajectory corresponding to $m$ counts at $t> t_m > \ldots > t_2 > t_1 > 0$ and no other photons in $(0,t]$ (called also an exclusive probability density). It is given by
\begin{eqnarray}
\lefteqn{ p_{0}^{t}(t_{m}, t_{m-1}, \ldots, t_{2}, t_{1}){dt_{m}dt_{m-1}\ldots dt_{1}} =}\\
&& ||\,|\alpha_{t|t_m,\ldots,t_1}\rangle \, ||^2 \int_{t}^{+\infty}dt^{\prime}|\xi_{t^{\prime}}|^2 +  ||\,|\beta_{t|t_m,\ldots,t_1}\rangle \, ||^2.\nonumber
\end{eqnarray} 
Probability of no counts up to time $t$
\begin{equation}\label{}
  P_0^t(0) = ||\,|\alpha_{t|\mathbf{0}_{t}}\rangle \, ||^2 \int_{t}^{+\infty}dt^{\prime}|\xi_{t^{\prime}}|^2+ ||\,|\beta_{t|\mathbf{0}_{t}}\rangle \, ||^2,
\end{equation}
which follows directly from (\ref{conprob}). The probability of having exactly $m$ counts up to time $t$ reads
\begin{equation}
P_{0}^{t}(m)\!=\!\int_{0}^{t}\!dt_{m}\!\int_{0}^{t_{m}}\!dt_{m-1}\!\ldots\!\int_{0}^{t_{2}}\!dt_{1}
p_{0}^{t}(t_{m}, \!t_{m-1}, \!\ldots \!, \!t_{2}, t_{1}) .
\end{equation}
To compute $P_0^t(0)$ one needs  conditional vectors $|\alpha_{t|\mathbf{0}_{t}}\rangle$ and $|\beta_{t|\mathbf{0}_{t}}\rangle$. Introducing
a non-Hermitian Hamiltonian (like in the Wigner-Weisskopf theory)
\begin{equation}
G=H_{S}-\frac{i}{2}L^{\dagger}L ,
\end{equation}
and the corresponding (non-unitary) propagator
\begin{equation}\label{}
  \mathbf{T}_t = e^{-iGt} ,
\end{equation}
one  finds
\begin{equation}\label{zeroa}
|\alpha_{t|\mathbf{0}_{t}}\rangle = \mathbf{T}_t |\psi\rangle
\end{equation}
from the limit
\begin{equation}\label{lim1}
\lim_{j\to +\infty}\left(\mathbbm{1}_{\mathcal{S}} - iG\frac{t}{j} + O\left(\frac{t^2}{j^2}\right)\right)^{j}|\psi\rangle\,
\end{equation}
and
\begin{eqnarray}\label{zerob}
|\beta_{t|\mathbf{0}_{t}}\rangle = -\int_{0}^{t}dt^{\prime} \mathbf{T}_{t-t^{\prime}}\xi_{t^{\prime}}L^{\dagger} \mathbf{T}_{t^{\prime}}|\psi\rangle
\end{eqnarray}
from
\begin{eqnarray}\label{lim2}
-\lim_{j\to\infty}\sum_{k=0}^{j-1}\frac{t}{j}\left(\mathbbm{1}_{\mathcal{S}} - i G \frac{t}{j}+ O\left(\frac{t^2}{j^2}\right)\right)^{j-k-1}\xi_{k}L^{\dagger}\nonumber\\
\times\left(\mathbbm{1}_{\mathcal{S}} - i G\frac{t}{j}+ O\left(\frac{t^2}{j^2}\right)\right)^{k}|\psi\rangle.
\end{eqnarray}
The limits (\ref{lim1}) and (\ref{lim2}) one obtains from (\ref{convec1a}) and (\ref{convec1b}) by taking (\ref{vmatrix}) and division $\tau=t/j$.

For a count at the time $t^{\prime}$ and no other counts in the interval $(0,t]$ we have the conditional vectors
\begin{equation}\label{veccon1}
|\alpha_{t|t^{\prime}}\rangle = \sqrt{dt^{\prime}}\mathbf{T}_{t-t^{\prime}} L \mathbf{T}_{t^{\prime}} |\psi\rangle ,
\end{equation}
and
\begin{eqnarray}\label{veccon2}
|\beta_{j|t^{\prime}}\rangle &=&  \sqrt{dt^{\prime}}\Big[  \mathbf{T}_t \xi_{t^{\prime}} - \mathbf{T}_{t-t'} L \left( \int_{0}^{t^{\prime}}ds \, \mathbf{T}_{t'-s} \xi_{s} L^{\dagger} \mathbf{T}_s \right) \nonumber \\
 &-&  \left(\int_{t^{\prime}}^{t} ds \mathbf{T}_{t-s} \xi_{s} L^{\dagger} \mathbf{T}_{s-t^{\prime}} \right) L \mathbf{T}_{t^{\prime}}
\Big]|\psi\rangle.
\end{eqnarray}
The conditional vector (\ref{veccon1}) one obtains as the limit
\begin{eqnarray}\label{lim3}
\lim_{j\to+\infty}\, \sqrt{t/j} \, \left(\mathbbm{1}_{\mathcal{S}} - i G \frac{t}{j}+ O\left(\frac{t^2}{j^2}\right)\right)^{j\left(1-\frac{t^{\prime}}{t}\right)}L\nonumber\\
\times\left(\mathbbm{1}_{\mathcal{S}} - i G\frac{t}{j}+ O\left(\frac{t^2}{j^2}\right)\right)^{\frac{jt^{\prime}}{t}-1}|\psi\rangle
\end{eqnarray}
and the conditional vector (\ref{veccon2}) from
\begin{widetext}
\begin{eqnarray}\label{lim4}
\lim_{j\to+\infty}\, \sqrt{t/j}\, \bigg[
\left(\mathbbm{1}_{\mathcal{S}} - i G \frac{t}{j}+ O\left(\frac{t^2}{j^2}\right)\right)^{j-1}\xi_{\frac{jt^{\prime}}{t}-1}\nonumber\\
-\left(\mathbbm{1}_{\mathcal{S}} - i G \frac{t}{j}+ O\left(\frac{t^2}{j^2}\right)\right)^{j\left(1-\frac{t^{\prime}}{t}\right)}
L\sum_{k=0}^{\frac{jt^{\prime}}{t}-2}\frac{t}{j}\left(\mathbbm{1}_{\mathcal{S}} - i G\frac{t}{j}+ O\left(\frac{t^2}{j^2}\right)\right)^{\frac{jt^{\prime}}{t}-2-k}\xi_{k}L^{\dagger}\left(\mathbbm{1}_{\mathcal{S}} - i G\frac{t}{j}+ O\left(\frac{t^2}{j^2}\right)\right)^{k}\nonumber\\
-\sum_{k=\frac{jt^{\prime}}{t}}^{j-1}\frac{t}{j}\left(\mathbbm{1}_{\mathcal{S}} - i G \frac{t}{j}+ O\left(\frac{t^2}{j^2}\right)\right)^{j-k-1}\xi_{k}L^{\dagger}
\left(\mathbbm{1}_{\mathcal{S}} - i G \frac{t}{j}+ O\left(\frac{t^2}{j^2}\right)\right)^{k-\frac{jt^{\prime}}{t}}L\left(\mathbbm{1}_{\mathcal{S}} - i G \frac{t}{j}+ O\left(\frac{t^2}{j^2}\right)\right)^{\frac{jt^{\prime}}{t}-1}\bigg]|\psi\rangle,
\end{eqnarray}
\end{widetext}
where we put $t^{\prime}=l_{1}t/j$ and by definition we have $dt=t/j$. The expressions one gets from (\ref{convec2a}) and (\ref{convec2b}).

In analogous way, for two counts at  $t^{\prime}$ and $t^{\prime\prime}$, where
$0<t^{\prime}<t^{\prime\prime}$, and no other counts in the interval $(0,t]$,
one gets
\begin{equation}
|\alpha_{t|t^{\prime \prime}, t^{\prime}}\rangle=\sqrt{dt^{\prime\prime}dt^{\prime}}
\mathbf{T}_{t-t^{\prime\prime}} L \mathbf{T}_{t^{\prime\prime}-t^{\prime}} L \mathbf{T}_{t^{\prime}}|\psi\rangle,
\end{equation}
\begin{eqnarray}
|\beta_{t|t^{\prime \prime}, t^{\prime}}\rangle &=&\sqrt{dt^{\prime\prime}dt^{\prime}}
\Big[
\mathbf{T}_{t-t^{\prime\prime}} L \mathbf{T}_{t^{\prime\prime}} \xi_{t^{\prime}} + \mathbf{T}_{t-t^{\prime}} \xi_{t^{\prime\prime}}L \mathbf{T}_{t^{\prime}} \nonumber\\
&-& \mathbf{T}_{t-t^{\prime\prime}} L \mathbf{T}_{t^{\prime\prime}-t^{\prime}} L \Big( \int_{0}^{t^{\prime}} ds \mathbf{T}_{t^{\prime}-s} \xi_{s} L^{\dagger} \mathbf{T}_s\Big)  \\
&-& \mathbf{T}_{t-t^{\prime\prime}} L \Big( \int_{t^{\prime}}^{t^{\prime\prime}}ds
\mathbf{T}_{t^{\prime\prime}-s} \xi_{s} L^{\dagger} \mathbf{T}_{s-t^{\prime}} \Big) L \mathbf{T}_{t^{\prime}}\nonumber\\
&-& \Big( \int_{t^{\prime\prime}}^{t}ds \mathbf{T}_{t-s} \xi_{s} L^{\dagger} \mathbf{T}_{s-t^{\prime\prime}} \Big)
L \mathbf{T}_{t^{\prime\prime}-t^{\prime}} L \mathbf{T}_{t^{\prime}} \Big]|\psi\rangle. \nonumber
\end{eqnarray}

At first sight  the above formulae seem to be complicated but the physical interpretation of individual terms are very intuitive. The term defined by the conditional vector $|\alpha_{t}\rangle$ gives the contribution to the probability of particular trajectory conditioned on the assumption that the two level system of the environment prepared in the upper state will appear after the time $t$, so all photons measured by us up to $t$ were emitted by the system $\mathcal{S}$. The conditional vector $|\beta_{t}\rangle$ gives a contribution to the probability based on the assumption that the system $\mathcal{S}$ has already interacted with the two level system of the environment prepared in the upper state.  So for instance, if we did not observe any photon up to the time $t$, it means the system $\mathcal{S}$ did not emit any photon and if it met photon of the external field in the period from $0$ to $t$ it absorbed it. For the case of some counts in the expressions for $|\beta_{t}\rangle$ one can recognize two sources of the measured photons: the external field and the system $\mathcal{S}$. The photon of the external field can be absorbed by $\mathcal{S}$ or directly measured.

The derived formulas can be applied for any quantum system $\mathcal{S}$. We show the solution to the problem taking as $\mathcal{S}$ a two-level atom prepared initially in the ground state $|0\rangle$. We take $L=\sqrt{\Gamma}\sigma^{-}$ and for simplicity $H_{\mathcal{S}}=0$. From (\ref{zeroa}) and (\ref{zerob}) one can easily find the analytical expression for the probability of not having any count in the time interval $(0, t]$
\begin{equation}
P_{0}^{t}(0)=\int_{t}^{+\infty}dt^{\prime}|\xi_{t^{\prime}}|^2+\Gamma e^{-\Gamma t}\left|\int_{0}^tdt^{\prime} \xi_{t^{\prime}}e^{\frac{\Gamma t^{\prime}}{2}}\right|^2.
\end{equation}
Moreover, we can check easily that the probabilities of two or more counts are equal to zero and the probability of having one count in the interval $(0,t]$ is equal to $1-P_{0}^{t}(0)$. By considering all possible trajectories one can check that the {\it a priori} solution has the form
\begin{equation}
(1-p(t))|0\rangle\langle 0|+p(t)|1\rangle\langle 1|
\end{equation}
where
\begin{equation}
p(t)=\Gamma e^{-\Gamma t}\left|\int_{0}^tdt^{\prime}\xi_{t^{\prime}} e^{\frac{\Gamma t^{\prime}}{2}}\right|^2
\end{equation}
is the {\it a priori} probability of excitation at time $t$.
The problem of the most efficient excitation of the two-level system by a single photon in a propagating mode was studied for instance in \cite{WMSS11, SAL10a, SAL10b}.  But instead of the numerical treatment of the problem, our approach offers the analogical formulas for the probabilities of particular trajectories.

\section{Conditional evolution for the diffusion process}

In this Section we consider the measurement of the observable
\begin{equation}
\sigma_{k}^{x}=
\sigma_{k}^{+}+\sigma_{k}^{-}=|+\rangle_{k}\langle+|-|-\rangle_{k}\langle-| ,
\end{equation}
with $k=0,1,2,\ldots$, and
\begin{eqnarray}\label{xbase}
|+\rangle_{k} &=&\frac{1}{\sqrt{2}}\left(|0\rangle_{k}+|1\rangle_{k}\right),\nonumber\\
|-\rangle_{k} &=&\frac{1}{\sqrt{2}}\left(|0\rangle_{k}-|1\rangle_{k}\right),
\end{eqnarray}
being vectors from the Hilbert space $\mathcal{H}_{\mathcal{E},k}$.

We prove that the conditional state of $\mathcal{S}$ and this part of the environment which has not interacted with $\mathcal{S}$ up to the time $j\tau$ at the moment $j\tau$ can be written in the form of (\ref{cond2}) with the conditional vectors $|\alpha_j\rangle$, $|\beta_j\rangle$ which satisfy the following recurrence equations
\begin{eqnarray}\label{rec2a}
    |\alpha_{j+1}\rangle &=& \frac{1}{\sqrt{2}}\left(V_{0 0} +q_{j+1} V_{10}\right)|\alpha_j\rangle,
\end{eqnarray}
\begin{eqnarray}\label{rec2b}
    |\beta_{j+1}\rangle &=& \frac{1}{\sqrt{2}}\left[\left(V_{0 0}+ q_{j+1} V_{10}\right) |\beta_j\rangle  \right.\nonumber \\ &+& \left. \sqrt{\tau}\xi_{j}\left(V_{01} + q_{j+1} V_{11}\right) |\alpha_j\rangle\right]
  \end{eqnarray}
with the initial condition $|\alpha_{0}\rangle=|\psi\rangle$, $|\beta_{0}\rangle=0$. By $q_{j+1}=1,-1$ we indicated the result of the measurement performed at $(j+1)\tau$ on the $j$-th qubits in the basis (\ref{xbase}).

The prove is straightforward  if we notice that Eq. (\ref{cond}) can be written in the form
\begin{widetext}
\begin{eqnarray}\label{der}
\mathbb{V}_{j}|\Psi_j\rangle&=&
\frac{1}{\sqrt{2}}|+\rangle_{j} \otimes\bigg\{\sum_{n=j+1}^{+\infty} \sqrt{\tau}\xi_{k}\sigma_{k}^{+} |vac \rangle_{[j+1,+\infty)} \otimes \left(V_{00}+V_{10}\right)|\alpha_j\rangle \nonumber\\
&+ & | vac \rangle_{[j+1,+\infty)} \otimes
\left(\left(V_{00}+V_{10}\right)|\beta_{j}\rangle
+\sqrt{\tau}\xi_{j}\left(V_{01}+V_{11}\right)|\alpha_j\rangle \right)\bigg\}\nonumber\\
&&+\frac{1}{\sqrt{2}}|-\rangle_{j}  \otimes\bigg\{\sum_{k=j+1}^{+\infty} \sqrt{\tau}\xi_{k} \sigma_{k}^{+}| vac \rangle_{[j+1,+\infty)} \otimes \left(V_{00}-V_{10}\right)|\alpha_j\rangle\nonumber\\
&+&| vac \rangle_{[j+1,+\infty)} \otimes
\left[ \left(V_{00}-V_{10}\right)|\beta_{j}\rangle+\sqrt{\tau}\xi_{j}
\left(V_{01}-V_{11}\right)|\alpha_j\rangle \right]\bigg\}.
\end{eqnarray}
\end{widetext}
The conditional vector $|\Psi_{j+1}\rangle$ from the Hilbert space   $\displaystyle{\bigotimes_{k=j+1}^{+\infty}}\mathcal{H}_{\mathcal{E},k}\otimes \mathcal{H}_{S}$ is in this case defined by
\begin{equation}
\left(\Pi_{+}\otimes \mathbbm{1}_{S}\right)\mathbb{V}_{j}|\Psi_j\rangle=|+\rangle_{j}\otimes|\Psi_{j+1}\rangle,
\end{equation}
\begin{equation}
\left(\Pi_{-}\otimes \mathbbm{1}_{S}\right)\mathbb{V}_{j}|\Psi_j\rangle=|-\rangle_{j}\otimes|\Psi_{j+1}\rangle,
\end{equation}
where $\Pi_{+}=|+\rangle_{j}\langle +|$ and $\Pi_{-}=|-\rangle_{j}\langle -|$. By (\ref{der}) we see that $|\Psi_{j+1}\rangle$ has the form
(\ref{cond1}) with the conditional vectors $|\alpha_{j+1}\rangle$, $|\beta_{j+1}\rangle$  given by (\ref{rec2a}) and (\ref{rec2b}) , and it ends the proof.

The form of $|\tilde{\Psi}_{j}\rangle$ reflects the fact the system $\mathcal{S}$ becomes entangled with the part of the environment which has not interacted with $\mathcal{S}$ yet. We get {\it a posteriori} state  of $\mathcal{S}$ by taking the trace from $|\tilde{\Psi}_{j}\rangle\langle\tilde{\Psi}_{j}|$
over the future environment space. We obtain the conditional normalized density matrix of the form (\ref{condS}) and for the conditional matrix $\rho_{j}$ we have

\begin{widetext}
\begin{eqnarray}
2\rho_{j+1} &=& V_{00} \rho_{j} V_{00}^\dagger + V_{10} \rho_{j} V_{10}^\dagger +\sqrt{\tau}\left[\xi_{j}^{\ast}\left(
V_{00} |\beta_{j}\rangle\langle\alpha_{j}|V_{01}^\dagger+
V_{10}|\beta_{j}\rangle\langle\alpha_{j}|V_{11}^\dagger\right)
+\xi_{j}\left(
V_{01} |\alpha_{j}\rangle\langle\beta_{j}|V_{00}^\dagger+
V_{11} |\alpha_{j}\rangle\langle\beta_{j}|V_{10}^\dagger\right)\right]\nonumber\\
&&+|\xi_{j}|^2\tau
\left(
V_{01} |\alpha_{j}\rangle\langle\alpha_{j}|V_{01}^\dagger +
V_{11} |\alpha_{j}\rangle\langle\alpha_{j}|V_{11}^\dagger -
V_{00}|\alpha_{j}\rangle\langle\alpha_{j}|V_{00}^\dagger -
V_{10} |\alpha_{j}\rangle\langle\alpha_{j}|V_{10}^\dagger\right)\nonumber\\
&&+q_{j+1}\left[V_{10} \rho_{j} V_{00}^\dagger + V_{00} \rho_{j} V_{10}^\dagger
++\sqrt{\tau}\xi_{j}^{\ast}\left(
V_{10} |\beta_{j}\rangle\langle\alpha_{j}|V_{01}^\dagger+
V_{00} |\beta_{j}\rangle\langle\alpha_{j}|V_{11}^\dagger\right)\right.\nonumber\\
&&\left.
+\sqrt{\tau}\xi_{j}\left(
V_{01} |\alpha_{j}\rangle\langle\beta_{j}|V_{10}^\dagger+
V_{11} |\alpha_{j}\rangle\langle\beta_{j}|V_{00}^\dagger\right)\right]\nonumber\\
&&+q_{j+1}|\xi_{j}|^2\tau
\left(
V_{01} |\alpha_{j}\rangle\langle\alpha_{j}|V_{11}^\dagger +
V_{11} |\alpha_{j}\rangle\langle\alpha_{j}|V_{01}^\dagger -
V_{10}|\alpha_{j}\rangle\langle\alpha_{j}|V_{00}^\dagger -
V_{00} |\alpha_{j}\rangle\langle\alpha_{j}|V_{10}^\dagger\right),
\end{eqnarray}
\begin{eqnarray}
2|\alpha_{j+1}\rangle\langle\beta_{j+1}| &=&  V_{00} |\alpha_{j}\rangle\langle\beta_{j}|V_{00}^\dagger+V_{10} |\alpha_{j}\rangle\langle\beta_{j}|V_{10}^\dagger
+\sqrt{\tau} \xi_j^*  \left(V_{00} |\alpha_{j}\rangle\langle\alpha_{j}|V_{01}^\dagger
+ V_{10} |\alpha_{j}\rangle\langle\alpha_{j}|V_{11}^\dagger\right)\nonumber\\
&&+q_{j+1} \left(V_{10} |\alpha_{j}\rangle\langle\beta_{j}|V_{00}^\dagger+
V_{00} |\alpha_{j}\rangle\langle\beta_{j}|V_{10}^\dagger\right)\nonumber\\
&&+q_{j+1}\sqrt{\tau} \xi_j^*  \left(V_{10} |\alpha_{j}\rangle\langle\alpha_{j}|V_{01}^\dagger+V_{00} |\alpha_{j}\rangle\langle\alpha_{j}|V_{11}^\dagger\right),\\
2|\alpha_{j+1}\rangle\langle\alpha_{j+1}| &=& V_{00} |\alpha_{j}\rangle\langle\alpha_{j}|V_{00}^\dagger+V_{10} |\alpha_{j}\rangle\langle\alpha_{j}|V_{10}^\dagger +q_{j+1}\left(V_{10} |\alpha_{j}\rangle\langle\alpha_{j}|V_{00}^\dagger+
V_{00} |\alpha_{j}\rangle\langle\alpha_{j}|V_{10}^\dagger\right).
\end{eqnarray}
\end{widetext}
Hence for a small $\tau$ and the matrix (\ref{vmatrix}) we obtain the difference equations 
\begin{widetext}
 \begin{eqnarray}\label{filter4}
2\rho_{j+1} &=& \rho_{j} -i[H_{S},\rho_{j}]\tau-\frac{1}{2}\{L^{\dagger}L,\rho_{j}\}\tau
+L\rho_{j}L^{\dagger}\tau 
+[L,|\beta_{j}\rangle\langle\alpha_{j}|]
\tau\xi_{j}^{\ast}+[|\alpha_{j}\rangle\langle\beta_{j}|,L^{\dagger}]\tau\xi_{j}\nonumber\\
&+& q_{j+1}\sqrt{\tau}\bigg( L\rho_{j}+\rho_{j}L^{\dagger}+|\beta_{j}\rangle\langle\alpha_{j}|\xi_{j}^{\ast}
+|\alpha_{j}\rangle\langle\beta_{j}|\xi_{j}\bigg) + O(\tau^2),
\end{eqnarray}

 \begin{eqnarray}
2|\alpha_{j+1}\rangle\langle\beta_{j+1}| &=&
|\alpha_{j}\rangle\langle\beta_{j}|-i[H_{S},|\alpha_{j}\rangle\langle\beta_{j}|]\tau
-\frac{1}{2}\{L^{\dagger}L,|\alpha_{j}\rangle\langle\beta_{j}|\}\tau 
+L|\alpha_{j}\rangle\langle\beta_{j}|L^{\dagger}\tau+[L,|\alpha_{j}\rangle\langle\alpha_{j}|]
\xi_{j}^{\ast}\tau \nonumber\\
&+&q_{j+1}\sqrt{\tau}\bigg(L|\alpha_{j}\rangle\langle\beta_{j}|+|\alpha_{j}\rangle\langle\beta_{j}|L^{\dagger}
+|\alpha_{j}\rangle\langle\alpha_{j}|\xi_{j}^{\ast}\bigg) + O(\tau^2),
\end{eqnarray}

 \begin{eqnarray}
2|\alpha_{j+1}\rangle\langle\alpha_{j+1}|&=&
|\alpha_{j}\rangle\langle\alpha_{j}|-i[H_{S},|\alpha_{j}\rangle\langle\alpha_{j}|]\tau
-\frac{1}{2}\{L^{\dagger}L,|\alpha_{j}\rangle\langle\alpha_{j}|\}\tau
+  L|\alpha_{j}\rangle\langle\alpha_{j}|L^{\dagger}\tau\nonumber\\
&+ &  q_{j+1}\sqrt{\tau}\bigg(L|\alpha_{j}\rangle\langle\alpha_{j}|+
|\alpha_{j}\rangle\langle\alpha_{j}|L^{\dagger}\bigg) + O(\tau^2).
\end{eqnarray}
\end{widetext}
The conditional probability of the outcome $q_{j+1}$ at the moment $(j+1)\tau$ if  the {\it a posteriori} state of $\mathcal{S}$ at $j\tau$ is
\begin{equation}
\tilde{\rho}_{j}=\frac{\rho_{j}}{\mathrm{Tr}\rho_{j}}
\end{equation}
is defined by
\begin{equation}
p_{j+1}(q_{j+1}|\tilde{\rho}_{j})=\frac{\mathrm{Tr}\rho_{j+1}}{\mathrm{Tr}\rho_{j}},
\end{equation}
where $\rho_{j+1}$ is given by (\ref{filter4}) and one can check that
\begin{equation}\label{prob}
p_{j+1}(q_{j+1}|\tilde{\rho}_{j})=\frac{1}{2}\left(1+q_{j+1}r_{j}\sqrt{\tau}\right)+O(\tau^2),
\end{equation}
where
\begin{equation}\label{intensity2}
r_{j}=\mathrm{Tr}\left(L\tilde{\rho}_{j}+\tilde{\rho}_{j}L^{\dagger}
+|\tilde{\beta}_{j}\rangle\langle\tilde{\alpha}_{j}|\xi_{j}^{\ast}+
|\tilde{\alpha}_{j}\rangle\langle\tilde{\beta}_{j}|\xi_{j}\right),
\end{equation}
where $|\tilde{\alpha}_{j}\rangle=|{\alpha}_{j}\rangle/\sqrt{\mathrm{Tr}\rho_{j}}$ and $|\tilde{\beta}_{j}\rangle=|{\beta}_{j}\rangle/\sqrt{\mathrm{Tr}\rho_{j}}$. Let us introduce the stochastic process
\begin{equation}
w_{j}=\sqrt{\tau}\sum_{k=1}^{j}\left(q_{k}-r_{k-1}\sqrt{\tau}\right).
\end{equation}
The process $w_{j}$ in the limit $\tau\rightarrow 0$ converges to the Wiener process. One can check using the formula (\ref{prob}) that the mean values $\mathbbm{E}[q_{k}|\tilde{\rho}_{k-1}]\simeq r_{k-1}\sqrt{\tau}$, $\mathbbm{E}[(q_{k+1})^2|\tilde{\rho}_{k-1}] \simeq 1$.
By making use of the following approximation for small values $\tau$ that
\begin{equation}
\frac{1}{\mathrm{Tr}\rho_{j+1}}\simeq\frac{2}{\mathrm{Tr}\rho_{j}}
\left(1-q_{j+1}r_{j}\sqrt{\tau}+r_{j}^2\tau\right)
\end{equation}
one may derive the set of difference stochastic equations
\begin{widetext}
 \begin{eqnarray}\label{filter5}
\tilde{\rho}_{j+1}-
\tilde{\rho}_{j}&=&-i[H_{S},\tilde{\rho}_{j}]\tau
-\frac{1}{2}\{L^{\dagger}L,\tilde{\rho}_{j}\}\tau+L\tilde{\rho}_{j}L^{\dagger}\tau 
+[|\tilde{\alpha}_{j}\rangle\langle\tilde{\beta}_{j}|,L^{\dagger}]\xi_{j}\tau+[L,|\tilde{\beta}_{j}\rangle\langle\tilde{\alpha}_{j}|]
\xi_{j}^{\ast}\tau\nonumber\\
&+& \bigg(L\tilde{\rho}_{j}+\tilde{\rho}_{j}L^{\dagger}+
|\tilde{\beta}_{j}\rangle\langle\tilde{\alpha}_{j}|\xi_{j}^{\ast}
+|\tilde{\alpha}_{j}\rangle\langle\tilde{\beta}_{j}|\xi_{j}
-\tilde{\rho}_{j}r_{j}\bigg)   \Delta w_{j+1}
\end{eqnarray}

 \begin{eqnarray}
|\tilde{\alpha}_{j+1}\rangle\langle\tilde{\beta}_{j+1}|&=&
|\tilde{\alpha}_{j}\rangle\langle\tilde{\beta}_{j}|-i[H_{S},|\tilde{\alpha}_{j}\rangle\langle\tilde{\beta}_{j}|]\tau
-\frac{1}{2}\{L^{\dagger}L,|\tilde{\alpha}_{j}\rangle\langle\tilde{\beta}_{j}|\}\tau 
+L|\tilde{\alpha}_{j}\rangle\langle\tilde{\beta}_{j}|L^{\dagger}\tau+[L,|\tilde{\alpha}_{j}\rangle\langle
\tilde{\alpha}_{j}|]
\xi_{j}^{\ast}\tau\nonumber\\
&+& \bigg(L|\tilde{\alpha}_{j}\rangle\langle\tilde{\beta}_{j}|+|\tilde{\alpha}_{j}\rangle\langle\tilde{\beta}_{j}|L^{\dagger}
+|\tilde{\alpha}_{j}\rangle\langle\tilde{\alpha}_{j}|\xi_{j}^{\ast}
-|\tilde{\alpha}_{j}\rangle\langle\tilde{\beta}_{j}|r_{j}\bigg)
 \Delta w_{j+1},
\end{eqnarray}

\begin{eqnarray}
|\tilde{\alpha}_{j+1}\rangle\langle\tilde{\alpha}_{j+1}|&=&
|\tilde{\alpha}_{j}\rangle\langle\tilde{\alpha}_{j}|
-i[H_{S},|\tilde{\alpha}_{j}\rangle\langle\tilde{\alpha}_{j}|]\tau 
- \frac{1}{2}\{L^{\dagger}L,|\tilde{\alpha}_{j}\rangle\langle\tilde{\alpha}_{j}|\}\tau
+L|\tilde{\alpha}_{j}\rangle\langle\tilde{\alpha}_{j}|L^{\dagger}\tau \nonumber\\
&+ &  \bigg(L|\tilde{\alpha}_{j}\rangle\langle\tilde{\alpha}_{j}|
+|\tilde{\alpha}_{j}\rangle\langle\tilde{\alpha}_{j}|L^{\dagger}
- |\tilde{\alpha}_{j}\rangle\langle\tilde{\alpha}_{j}|r_{j}\bigg) \Delta w_{j+1} ,
\end{eqnarray}
\end{widetext}
where
$$  \Delta w_{j+1}  = w_{j+1} - w_j = q_{j+1}\sqrt{\tau}-r_{j}\tau . $$

In the limit $\tau\rightarrow 0$ we obtain the following stochastic differential equations of the form
\begin{widetext}
 \begin{eqnarray}
d\tilde{\rho}_{t}&=&-i[H_{S},\tilde{\rho}_{t}]dt-
\frac{1}{2}\left\{L^{\dagger}L,\tilde{\rho}_{t}\right\}dt
+L\tilde{\rho}_{t}L^{\dagger}dt  +[\tilde{\rho}^{01}_{t},L^{\dagger}]\xi_{t}dt+[L, \tilde{\rho}^{10}_{t}]\xi^{\ast}_{t}dt\nonumber\\
&+ &  \left(
L\tilde{\rho}_{t}+ \tilde{\rho}_{t}L^{\dagger}+\tilde{\rho}^{01}_{t}\xi_{t} +\tilde{\rho}^{10}_{t}\xi^{\ast}_{t}-
\tilde{\rho}_{t}r_{t}\right)dw(t),
\end{eqnarray}
\begin{eqnarray}
d\tilde{\rho}^{01}_{t}&=& -i[H_{S},\tilde{\rho}^{01}_{t}]dt-\frac{1}{2}\left\{L^{\dagger}L,\tilde{\rho}^{01}_{t}
\right\}dt+L\tilde{\rho}^{01}_{t}L^{\dagger}dt +\left[L,\tilde{\rho}^{00}_{t}\right]\xi^{\ast}_{t}dt\nonumber\\
&+& \left(L\tilde{\rho}^{01}_{t}+
\tilde{\rho}^{01}_{t}L^{\dagger}+\tilde{\rho}^{00}_{t}
\xi^{\ast}_{t}-\tilde{\rho}^{01}_{t}r_{t}\right)dw(t),
\end{eqnarray}
\begin{eqnarray}
d\tilde{\rho}^{00}_{t}&=&-i[H_{S},\tilde{\rho}^{00}_{t}]dt
-\frac{1}{2}\left\{L^{\dagger}L,\tilde{\rho}^{00}_{t}
\right\}dt+L\tilde{\rho}^{00}_{t}L^{\dagger}dt +\left(L\tilde{\rho}^{00}_{t}+
\tilde{\rho}^{00}_{t}L^{\dagger}-\tilde{\rho}^{00}_{t}r_{t}\right)dw(t),
\end{eqnarray}
\end{widetext}
where 
\begin{equation}
r_{t}=\mathrm{Tr}\left(L\tilde{\rho}_{t}+\tilde{\rho}_{t}L^{\dagger}
+\tilde{\rho}^{10}_{t}\xi_{t}^{\ast}+\tilde{\rho}^{01}_{t}
\xi_{t}\right),
\end{equation}
$\tilde{\rho}^{10}_{t}=(\tilde{\rho}^{01}_{t})^{\dagger}$, and initially we have $\tilde{\rho}_{0}=|\psi\rangle\langle\psi|$,
$\tilde{\rho}^{01}_{0}= 0$, and $\tilde{\rho}^{00}_{0}=|\psi\rangle\langle\psi|$.

\section{Conclusions}

In this paper we derived a filtering equation (quantum trajectories) for a system interacting with the environment prepared in a continuous mode single photon state. The initial state of `system + photon field' is factorized, however, the initial state of the field is highly entangled (depending on the photon profile $\xi_t$). We consider both the counting and diffusion processes. Although such filtering equation was already derived by Gough et al \cite{GJN11,GJNC12a} our approach is different and much simpler. Authors of \cite{GJN11, GJNC12a} start with the general quantum stochastic differential equation of Hudson and Parthasarathy \cite{HP84} and apply the general technique to the case of correlated single photon state. It is, therefore, clear that one has to assume the reader is familiar with quantum stochastic calculus. Our approach is more direct: starting with the simple repeated interaction model \cite{AP06, P08, PP09, P10} we present the model of a quantum system interacting with an infinite chain of identical and independent quantum systems representing the environment. This approach gives an intuitive and rigorous interpretation for the conditional evolution of the open quantum and quantum trajectories. Another advantage of our approach is simple and very intuitive interpretation of quantum trajectories. In particular we derive probability density of observing a particular trajectory corresponding to $m$ counts at $t> t_m > \ldots > t_2 > t_1 > 0$ and probability of no counts up to time $t$. Our filtering equations in a continuous limit are consistent with the results given in \cite{GJN11, GJNC12a, BC17}. It should be stressed that initial correlations of the photon field imply that the averaged filtering equation (over all possible results of measurements) is not of the standard Kossakowski--Lindbald form. Due to the presence of initial correlations the resulting averaged dynamics is highly non-Markovian. 
Interestingly, the non-Markovian nature of the physics here is rather due to the pre-existing entanglement in the environment, and not to any information back flow as is often the case in other non-Markovian systems \cite{BLP}. This problem we are going to analyze in the forthcoming paper.

\section*{Acknowledgements}

We thank anonymous referee for valuable comments.
This paper was partially supported by the National Science Center project 2015/17/B/ST2/02026.

\end{document}